\documentclass [pre,twocolumn]{revtex4}
\usepackage{graphicx}

\begin{document}                                                                
                                                                                
\title{Analytic treatment of a trading market model} 

\author{Arnab Das}

\author{Sudhakar Yarlagadda}

\affiliation{Saha Institute of Nuclear Physics, 1/AF Bidhan Nagar,
 Kolkata 700 064, India}             

\begin{abstract}
We mathematically analyze a simple market model where trading at each point in
 time involves only two agents with the sum of their money being conserved
and with neither parties resulting with negative money after the interaction
 process. The exchange involves random re-distribution among the two players
 of a fixed fraction of their total  money. We obtain a simple
integral nonlinear equation for the money distribution.
 We find that the zero savings and 
finite savings cases belong to different universality classes. 
While the zero savings case can be solved analytically, the finite
savings solution is obtained by numerically solving the integral equation.
 We find remarkable agreement with results obtained by other researchers using
sophisticated numerical techniques \cite{chat}.
\end{abstract}

\maketitle
                                                                                
{ The distribution of wealth in a nation among its people has been
of considerable interest over the ages. If
 one succeeds in mathematically
modeling the distribution it has serious implications. 
 One of the empirical laws known is that due to Pareto \cite{pareto}
 which states that
 the higher income group distribution decays like a power law with exponent
 between $2$ and $3$. On the other hand the lower income group distribution 
is Gibb's like. The Gibb's law has been numerically
demonstrated to be obtainable when trading 
between two agents is completely random and does not involve
any savings \cite{chak,chatchak,drag}. Also the finite savings case has been
studied numerically by Chakraborti and Chakrabarti \cite{chatchak}.
However an analytic treatment of these models is lacking and hence the present
work is aimed at meeting this need.

In this paper we derive analytically an integral equation for the 
probability distribution function of the money in the system.
The model describes trading between two agents $i$ and $j$ each
with money $m_{i}(t)$ and $m_{j}(t)$ at instant $t$.
Each agent saves a fraction $\lambda$ of his or her money
and the rest is traded. Trading involves a random
 re-distribution of the money available for trading 
[$(1-\lambda )(m_{i}(t) + m_{j} (t))$] among the two agents $i$ and $j$.
After trading , i.e., at instant $t+1$, 
$m_{i} (t+1)=m_{i} (t) + \Delta m$ and 
$m_{j} (t+1)=m_{j} (t) - \Delta m$ where
$\Delta m = [\epsilon (m_i + m_j) - m_{i} (t)](1-\lambda )$
with $0 \leq \lambda \leq 1$ and $\epsilon$ being a random number
between $0$ and $1$.

We will now derive the equilibrium distribution function
 $f(y)dy$ which gives the 
probability of an agent having money between $y$ and $y+dy$. 
We assume that, irrespective of what the starting point is, the
system evolves to the equilibrium distribution after sufficiently
long time. We will now consider interactions after the
system has attained steady state.
The joint probability that, before interaction, money 
of $i$ lies between $x$ and $x+dx$ and money of $j$ lies between $z$ and $z+dz$
is $f(x)dxf(z)dz$. Since the total money is conserved in the interaction,
we let $L=x+z$ and analyze in terms of $L$.
Then the joint probability becomes
 $f(x)dxf(L-x)dL$. We will now generate equilibrium distribution
after interaction by noting that
{\it at steady state the distribution is the same before and after
interaction}. Probability that $L$ is distributed to give money of $i$ between
$y$ and $y+dy$ is 
 \begin{equation} 
\frac{dy}{(1-\lambda ) L}
 f(x)dxf(L-x)dL ,
 \end{equation} 
 with $x \lambda \le y \le x \lambda
+ (1 - \lambda ) L$. Thus we see that  $x \leq y/\lambda$  and 
$x \geq [y- (1-\lambda )L]/\lambda$. Actually $x$ should also satisfy
the constraint  $0\le x \le L$ because the agents cannot have negative money.
Thus the upper limit on $x$ is $\min \{L,y/\lambda\}$ (i.e., minimum
of $L$ and $y/\lambda $)  and the lower limit
is $\max \{0,[y- (1-\lambda )L]/\lambda\}$.
Now, we know that the total money $L$ has to be greater than $y$ so that
the agents have non-negative money.
Thus we get the following distribution function
for the money of $i$ to lie between $y$ and $y+dy$
\begin{eqnarray} 
f(y)dy & = & dy\int_{y}^{\infty} \frac{dL}{(1 - \lambda ) L} \nonumber \\
 & & \int_{\max [0, \{y-(1-\lambda)L \} 
/\lambda]}^{\min [L,y/\lambda ]}
dx f(x)f(L-x) .
\label{distlamb}
\end{eqnarray}

It is of interest to note that when $\lambda = 0$, the upper
and lower limits of the $x$ integration become $0$ and $L$
and we get
 \begin{equation} 
f(y)dy=dy\int_{y}^{\infty} \frac{dL}{L}
\int_{0}^{L}
dx f(x)f(L-x) .
\label{distgibbs}
 \end{equation} 
We will now provide an analytic solution for the above zero savings case.
We first note that the double derivative with respect to $y$ of the
above Eq. (\ref{distgibbs}) yields
 \begin{equation} 
f^{ \prime}(y)
+y f^{\prime \prime}(y) = -f(y) f(0) - \int _{0}^{y} f(x)
 f^{\prime} (y-x) dx .
\label{distderiv}
 \end{equation} 
For small $y$ we assume that the function $f(y)$ and its first
and second derivatives are 
well behaved. Then as $y \rightarrow 0$ we get
 \begin{equation} 
f^{ \prime}(y)
\approx -f(y) f(0) .
\label{distapp}
 \end{equation} 
Then the solution for small $y$, after using the constraint $\int _{0}^{\infty}
f(y) dy = 1 $, is given by
 \begin{equation} 
f(y)
\approx f(0) exp {[-y f(0)]} .
\label{distappsol}
 \end{equation} 
Next we also note that the above function (in Eq. (\ref{distappsol}))
also solves the parent Eq. (\ref{distgibbs}) {\it exactly}. It
is of interest to note that there are similarities between
this distribution and the classical 
$Maxwell-Boltzmann$ distribution with the maximum corresponding
to $y=0$.

  We will now make a few observations regarding $f(y)$. On examination
of Eq. (\ref{distlamb}), assuming $f(y) \rightarrow 0$ 
as $y \rightarrow \infty $ and that $f(y) $ is well behaved
as $y \rightarrow 0 $, it is clear that
$\lim _{y \rightarrow 0} f(y) \rightarrow 0$.
Physically this makes sense because if everyone has non-zero savings, then 
a person with zero money will,  due to interactions, tend to non-zero
money faster than returning to zero money. A person with zero
money after a single interaction has  probability $1$
of getting non-zero money.  Also, once a person
has  non-zero money it takes an infinite number of interactions for that
person to end up with zero money. Thus the $\lambda =0 $ case and 
$\lambda > 0$ case belong to different universality classes.
Moreover $\lambda =1$ also belongs to a different universality because it
 is a static situation. Also it can be shown that for $0 < \lambda < 1$
as $y \rightarrow \infty$ the function decays exponentially.

 We have solved the integral equation given by Eq. (\ref{distlamb})
for the non-trivial case of $f(y) \neq \delta (y)$.
Our approach is based on solving for $f(y)$ iteratively by starting with
a trial function. To achieve convergence, we use a novel approach,
wherein as soon as $f(y)$ is calculated at any  $y$ it is immediately
used to evaluate $f(y)$ at other values of $y$. Thus we do not need to wait till
$f(y)$ is evaluated completely for all values of $y$ to use it
to get the next approximation for $f(y)$ as one might naively  expect. 
Details of this
iterative procedure and a critical analysis will be presented elsewhere.
We give here results (see Fig. 1) for integrations
 done with rectangular boxes of
width $0.01$ and upper limit of $L=10$.   All the  curves are
normalized and correspond to
 average money being {\it unity}. As $\lambda $ increases,
 for $\lambda > 0$,
 the peak value of the bell-shaped curves increases and shifts to
 the right and also the width decreases. It is of interest to note the 
rather striking similarity between our plots and those reported by 
A. Chatterjee, B. K. Chakrabarti, and S.S. Manna \cite{chat}.  It is 
very gratifying to note that by using completely different approaches
 one gets very similar results which lends confidence to these
approaches.

 In future we hope to extend our treatment to obtain a model
that is consistent with Pareto law. Numerical advances in
obtaining the Pareto law have been reported in Ref. \cite {condmat}.

   The authors are very grateful to B. K. Chakrabarti for inspiring
and useful discussions. One of the authors (S. Y.) also thanks
R. Ramakumar for support and helpful discussions.

\vspace{0.3cm}
{\centering \resizebox*{8cm}{6.5cm}{\rotatebox{270}{\includegraphics{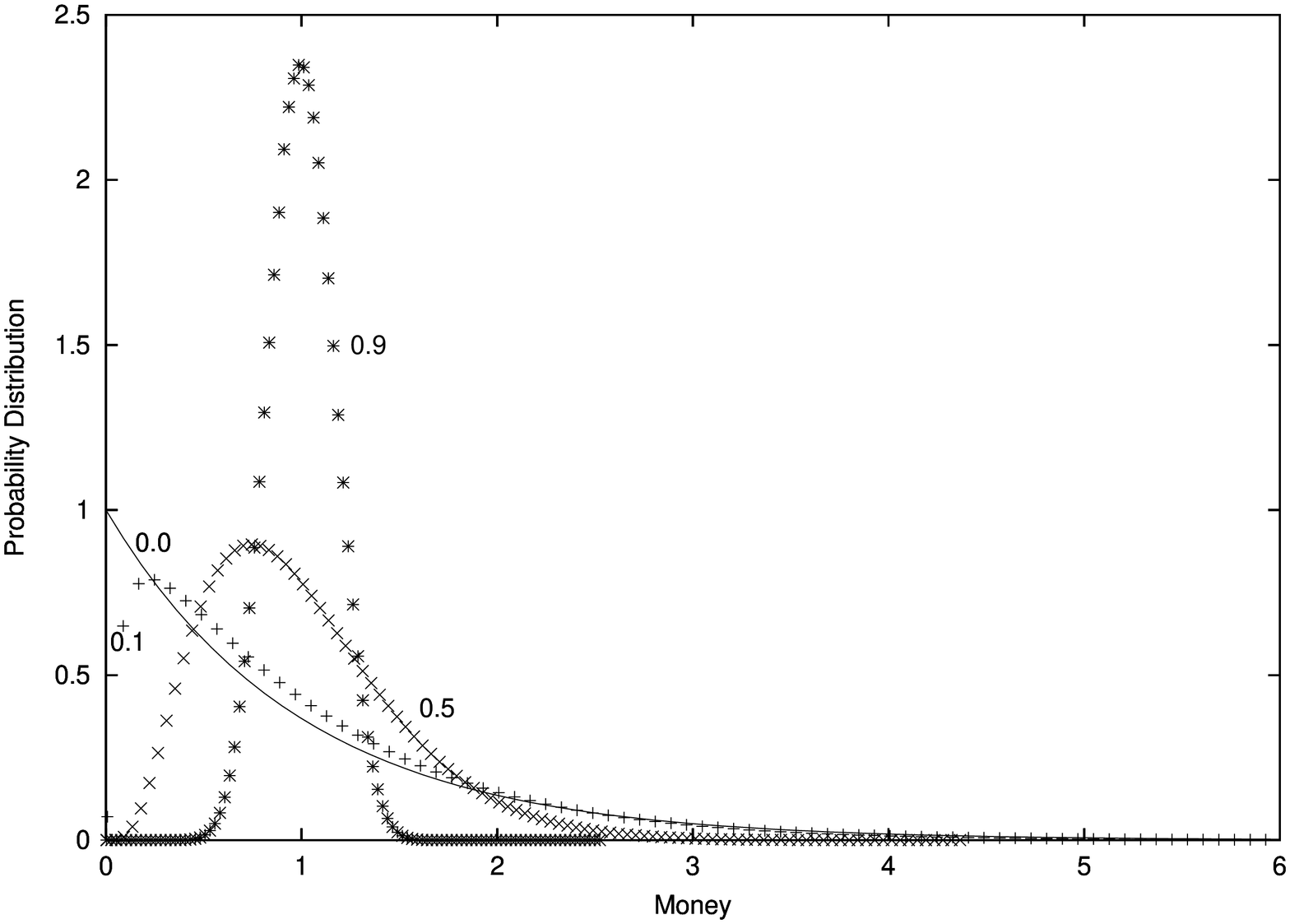}} \par}}
\vspace{0.3cm}
\vskip 0.1in
\noindent {\footnotesize FIG. 1: Plot of the money probability distribution 
function for various savings values ($\lambda = 0.0, 0.1, 0.5, 0.9$). 
The average money per person is set to unity.}{\footnotesize \par}

\end{document}